# Analysis of Relation between Motor Activity and Imaginary EEG Records

Enver Kaan Alptürk[1*], Yakup Kutlu[1]

[1]Iskenderun Technical University, Institute of Engineering and Natural Sciences, Hatay, Turkey

**Abstract**

Electroencephalography (EEG) signals signals are often used to learn about brain structure and to learn what thinking. EEG signals can be easily affected by external factors. For this reason, they should be applied various pre-process during their analysis. In this study, it is used the EEG signals received from 109 subjects when opening and closing their right or left fists and performing hand and foot movements and imagining the same movements. The relationship between motor activities and imaginary of that motor activities were investigated. Algorithms with high performance rates have been used for feature extraction , selection and classification using the nearest neighbour algorithm.

## 1. INTRODUCTION

Electroencephalography (EEG) signals are biological electrical signals obtained from the brain. EEG signals are affected by the electrical and magnetic field in the environment and all physical and mental activities of the individual receiving the signal. Since the first signal received by Hans Berger (Berger, 1929) from a human brain in the 1920s, EEG technology has continued to evolve in remote communication, brain-computer interfaces, and more (Kamarajan, 2019).... EEG signals are low amplitude and low frequency signals. The amplitudes of the significant signals vary between 0-200 mV. The frequency range varies between 0-100 Hz. If the level of activity in the brain increases, the frequency of EEG signals increases (Gür, Kaya, & Türk, 2014).

According to the frequencies and amplitudes of the EEG signals, information about the activities in the brain can be obtained. The delta waves with the lowest frequency are between 0-3.5 Hz and appear in deep sleep in adult humans. Where consciousness is closed, delta waves are produced intensely. Theta waves are between 4-8 Hz. and mostly occur in stressful situations. Alpha waves are the bridge between deep sleep and normal life and have a frequency between 8-13 Hz. Alpha waves intensify during alertness, during the transition to listening mode and relaxation. It is seen in situations where awareness and comprehension are high. Beta waves vary between 13-30 Hz. and are affected by all mental activities. It occurs in cases of concentrating, active thinking, and solving daily problems (Shaker, 2006).

---

[*] *Corresponding Author: Enver Kaan Alptürk, E-mail: kaanalpturk@outlook.com*

All of the physical and mental activities are carried out by the relevant lobe in the brain. Physical movements known as motor activity are processed in the frontal lobe. Other functions of the frontal lobe; reasoning, problem solving, decision making, controlling feelings and directing attention (Avci, Eeduran, & Yağbasan., 2014). EEG signals can be received by placing electrodes at various points in the frontal region where the frontal lobe is located, and the received signals provide information about physical and mental activities. Therefore, when using the frontal lobe actively, condensation is observed in alpha waves (Fink, Andreas, & al., 2018). In specific studies, the electrodes to be used must be placed in the area where the lobe where the activity takes place. Since the brain has a large surface and too many electrodes are used to find the right signal. Reducing the number of electrodes will both reduce the number of data and speed up the inspection process. Also, the fewer electrodes are found, the greater the portability of the device from which EEG signals are received.

In 2014, Gür et al., passed the moving averages filter as a low-pass filter with epileptic EEG signals and normal EEG signals, and derivative-based filters as a high-pass filter and applied FFT by removing signals from noise. In his findings, he found that the signals received from epileptic subjects in the range of 0-10 Hz differed significantly from those received from normal people (Gür, Kaya, & Türk, 2014).

Yağanoğlu et al., examined the feature extraction methods that can be used to analyze and classify EEG signals in 2014. They compared total, average, variance, standard deviation, polynomial fitting and similar methods by classifying them with the closest neighbor algorithm and achieved the highest performance rate with 87.3% (Yaganoglu, et al., 2014).

In this study, the frequency spectrum of the signals was extracted by applying FFT to EEG signals received from the 18 electrodes placed in the frontal lobe of the brain for 30 seconds at 160 Hz sampling frequency during the opening and closing right and left fists, hand and foot movements, and imagining that actions of 109 volunteers. PhysioNet EEG database recorded with BCI2000 instrumentation system was used (Schalk, McFarland, Hinterberger, Birbaumer, & Wolpaw, 2004) (Goldberger, et al., 2000). Feature extraction were made by various methods from the data resulting from the application of FFT. At the same time, the polynomial fitting method, the best method mentioned in (Yaganoglu, et al., 2014), was used to extract the features of the EEG signals by a different method. After the feature extraction, the signals of the individuals who performed the actions and those who dreamed of the actions were classified using the closest neighbor algorithm and the channels in which the highest performance was obtained were specified.

## 2. MATERIALS AND METHODS

In this study, some of the feature extraction methods with high performance rates in the literature are used to classify and match the motor activity and imagination actions in the dataset. Figure 1 shows the general flow diagram and the extracted features with the methods used.

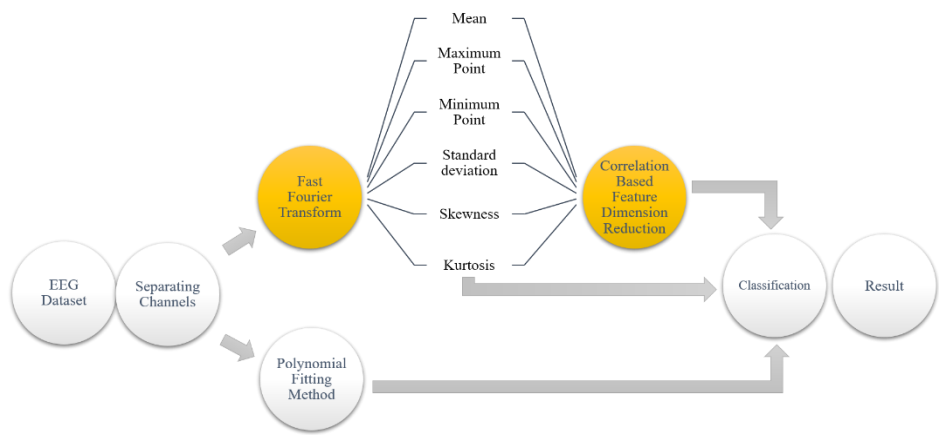

**Figure 1.** General Flow Chart and Application

### 2.1 EEG Data

EEG signals consisted of 3 repetitions of one minute from 109 volunteers during punch open and close movements, moving hands and feet at the same time, imagining punch open and close movements and imagining that they move their hands and feet at the same time, were recorded with 60 channels of BCI2000 instrumentation system. In this study, EEG signals were reduced to 30 seconds, and the signals received during punch open and close movements, moving hands and feet at the same time used as training data; and signals recorded during imagination of punch open and close movements and imagination that they move their hands and feet at the same time used as test data.

The intensity of the EEG signals varies depending on the sampling frequency and the number of channels. Working with raw EEG data is troublesome and useless. Therefore, EEG signals are subjected to a number of pretreatments before being classified or analyzed. In this study, some of the methods with the highest performance in EEG signals in previous studies have been applied to "Fc5", "Fc3", "Fc1", "Fcz", "Fc2", "Fc4", "Fc6", "F7", "F5", "F3", "F1", "Fz", "F2", "F4", "F6", "F8", "Ft7", "Ft8" channels in the frontal lobe in the data set used and the performance rates obtained by the nearest neighbor algorithm are compared. The channels used in this project from the electrodes placed on the skull in Figure 2 are shown in red.

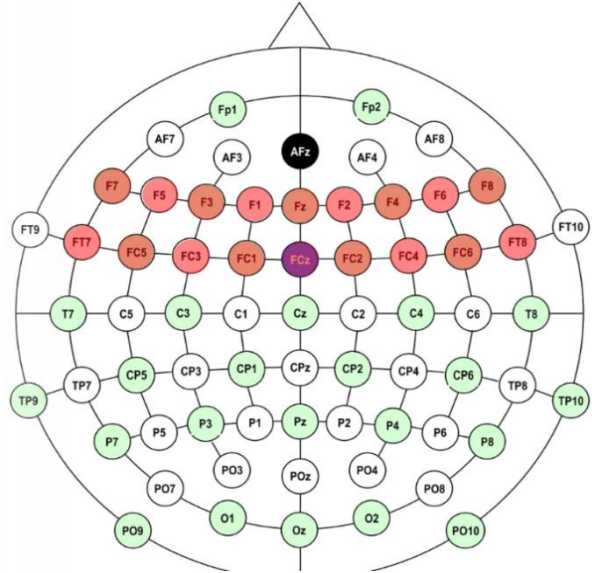

**Figure 2.** Display of the Channels Used on the Skull

### 2.2 Feature Extraction

The first step of the other methods applied to the data set is FFT. The FFT correlation used for extracting the frequency spectrum of EEG signals is shown in (Eq 2). The specified $X_k$ value represents the FFT coefficient. $N$ represents the total number of samples and $x$ represents the sampled signal data (Murugappan & Murugappan, 2013).

$$X_k = \sum_{n=0}^{N-1} x(n)^{-i\left(2\pi \times k \times \frac{n}{N}\right)} \qquad \text{(Eq. 2)}$$

For each 5 Hz part of 80 Hz data obtained after applying FFT, 6 features were extracted. These features are; mean, standard deviation, maximum point, minimum point, skewness and kurtosis. In all 80 Hz, 96 features were created for each channel. High performance values were obtained in the study in (Yaganoglu, et al., 2014) with some of the specified features.

In the next step, a distinction is made between the features. The feature separation was made by correlation calculation.

### 2.3 Feature Selection and Classification

The number of features formed in this method is as much as the order value. After the features were obtained, correlation based feature selection algorithm is used to discard unnecessary features. Correlation coefficients between variables are calculated. This shows the relationship between the two variables. It is the main diagonal line going from top left to bottom right, showing that each variable is always perfectly related to it. Therefore we have motor movements and imagine of the same movements features. Analyzing correlation of two variables suitable features could be determined

The motor movements were used as training data, and the imagine of the same movements were used as test data, and was classified using K Nearest Neighbor (KNN) algorithm. KNN is one of the most popular algorithm in machine learning models. It is stable and almost has no parameters according to other algorithms. K is used as 1 and Euclid distance used as distance parameter of KNN

## 3. RESULTS

When the EEG signals whose frequency spectrum is extracted with FFT are examined, it was seen that the subjects produced strong alpha waves while performing or imagining the actions. At the same time, attenuation has been detected in low beta waves. In Figure 3, the frequency spectrum of the signals belonging to the two groups and the powers of alpha and low beta waves are shown.

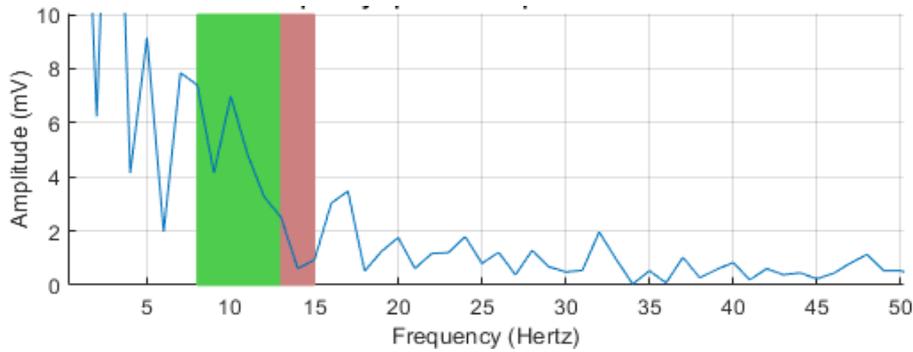

**Figure 3.** Power of Frequency Spectrum and Alpha and Low Beta Waves

The green region in Figure 3 shows the frequency range of alpha waves and the red region shows the frequency range of low beta waves. The rise in medium beta waves and alpha waves suggest that mental and physical activities are increasing.

At this stage, the average, standard deviation, maximum point, minimum point, skew and kurtosis features extracted for each 5 Hz. part of the 80 Hz. data obtained by applying FFT to the dataset are classified directly by the nearest neighbor algorithm and the classification performance according to the channels is shown in Figure 3.

Based on the performance percentages shown in Figure 3, it is concluded that the features obtained from FFT are not sufficient for classification alone. For this reason, in order to select the more meaningful features that high performance can be obtained, strong features have been determined by calculating correlations with the features of imagination and motor movements. In this study, correlation calculation shows the level of correlation between motor movements and imagination.

Features with a correlation level higher than 0.7 were chosen as strong features, and these features were used in the training and test data in the next classification. In Table 1, the correlation matrix created for the first 10 of the features created after the FFT application of the first EEG signals received from the Ft8 channel during the punch open and close movements and imagining actions of a randomly selected volunteer is shown.

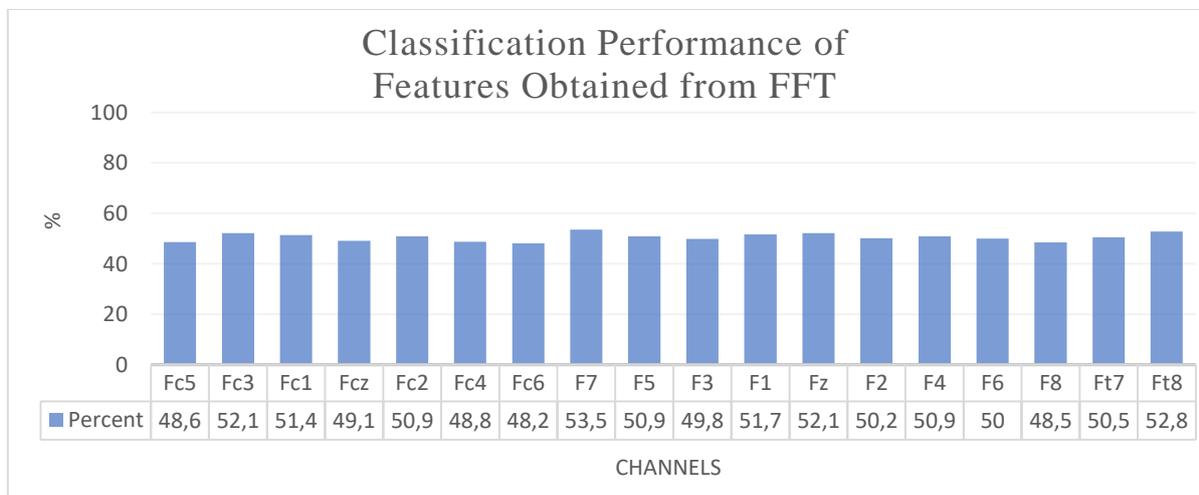

**Figure 3.** Classification Performance of Features Obtained from FFT

**Table 1.** Correlation matrix for the first 10 of the features between motor movement and imagination data,

| Imagine / Movement | 1 | 2 | 3 | 4 | 5 | 6 | 7 | 8 | 9 | 10 |
|---|---|---|---|---|---|---|---|---|---|---|
| 1 | **0,88784** | 0,860214 | 0,7209 | 0,82767 | -0,07965 | -0,18034 | 0,662731 | 0,686651 | 0,602852 | 0,669337 |
| 2 | 0,84866 | **0,870501** | 0,652344 | 0,860687 | 0,036497 | -0,05338 | 0,521474 | 0,549433 | 0,500567 | 0,525817 |
| 3 | 0,73111 | 0,674331 | **0,622896** | 0,627084 | -0,1342 | -0,23496 | 0,695202 | 0,695067 | 0,606094 | 0,694135 |
| 4 | 0,81194 | 0,849008 | 0,612882 | **0,846947** | 0,070516 | -0,00875 | 0,44824 | 0,47824 | 0,445378 | 0,45208 |
| 5 | 0,04852 | 0,096181 | -0,17925 | 0,15767 | **0,489568** | 0,526607 | -0,31585 | -0,31336 | -0,21386 | -0,33515 |
| 6 | 0,12118 | 0,006656 | -0,18052 | 0,058125 | 0,47227 | **0,478644** | -0,29635 | -0,30994 | -0,22473 | -0,32635 |
| 7 | 0,64883 | 0,533619 | 0,630939 | 0,468812 | -0,23087 | -0,36124 | **0,86082** | 0,888732 | 0,735183 | 0,88725 |
| 8 | 0,65048 | 0,543664 | 0,612996 | 0,484523 | -0,24112 | -0,34399 | 0,85825 | **0,88256** | 0,764179 | 0,869922 |
| 9 | 0,52294 | 0,429037 | 0,499352 | 0,378464 | -0,18283 | -0,32771 | 0,674765 | 0,694462 | **0,528964** | 0,718701 |
| 10 | 0,61051 | 0,511663 | 0,567194 | 0,457861 | -0,2326 | -0,30613 | 0,82053 | 0,84275 | 0,761134 | **0,817193** |

After the correlation is applied to the 96-feature motor movement and imagination data, a different number of strong features are determined for each channel. Features with a value higher than 0.7 as in Table 1, 41 strong features were selected for the "Ft8" channel according to correlation matrix.

The same procedure was applied on all of its data, and the showed that the highest number of strong features was 41 and the lowest number of strong features was 35 seen on result of correlation calculation. Looking at the average of all channels, 37 powerful features were found per channel.

As a result of this process, new features are again classified with the nearest neighbor algorithm and performance results according to the channels are shown in Figure 4.

It is seen in Figure 4 that the percentage of performance has increased significantly with the use of strong features after the correlation calculation. Especially the performance seen in "Fc1", "Fcz", "Fc2", "F6" and "Ft8" channels is satisfactory.

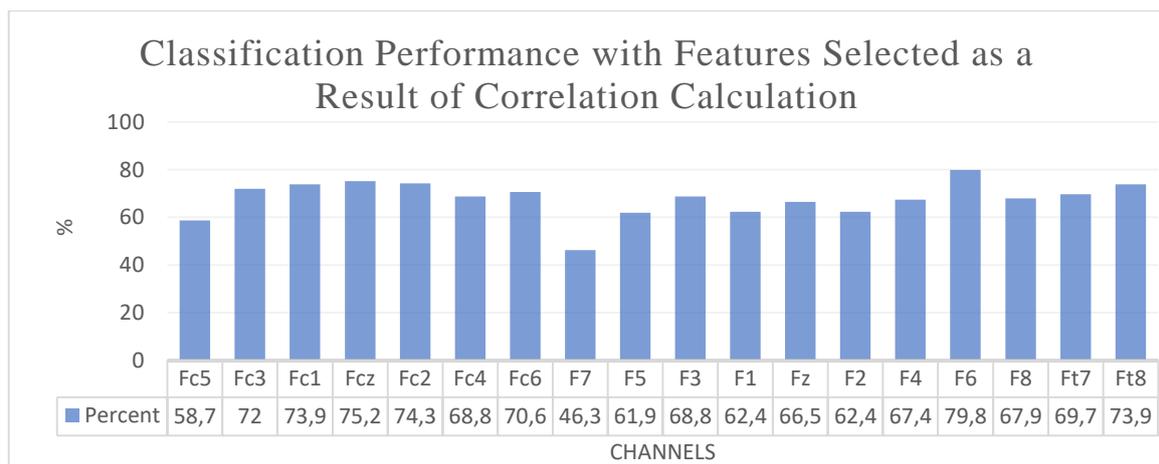

**Figure 4.** Classification Performance with Features Selected as a Result of Correlation Calculation

## 4. CONCLUSION AND DISCUSSION

It was concluded that the algorithm, feature extraction, selection and classification of the data set, has high performance rate. Frequency analysis was performed by applying FFT to the dataset and low performance percentages occurred at first because the feature set contained meaningless data. The correlation based feature reduction algorithm was achieved to select strong features and as a result of this algorithm, the performance increased significantly. It has been observed that EEG signals received from "Fc1", "Fcz", "Fc2", "F6" and "Ft8" channels produce more efficient results during motor movements. In addition, it has been observed that there are increases in alpha and middle beta waves and they are common in the signals received during the realization and imagination of the punch open and close movements. In future studies, an EEG device with increased portability can be designed by using only channels with successful results. In this way, devices that facilitate the use of physical movements in prostheses will be produced.

5. **REFERENES**